\documentclass[iop,apj,numberedappendix]{emulateapj}

\usepackage{natbib}
\usepackage{graphicx}
\usepackage{epstopdf}
\usepackage{amsmath,amsfonts}
\slugcomment{Submitted to ApJ, October 11, 2012}

\def\lsim{\mathrel{\raise.3ex\hbox{$<$\kern-.75em\lower1ex\hbox{$\sim$}}}}
\def\gsim{\mathrel{\raise.3ex\hbox{$>$\kern-.75em\lower1ex\hbox{$\sim$}}}}
\def\gtwid{\mathrel{\raise.3ex\hbox{$>$\kern-.75em\lower1ex\hbox{$\sim$}}}}
\def\proptwid{\mathrel{\raise.3ex\hbox{$\propto$\kern-.75em\lower1ex\hbox{$\sim$}}}}

\defcitealias{tms}{TMS}
\setcitestyle{notesep={; }}

\begin{document}
\title{Optimal Correlation Estimators for Quantized Signals}
\shorttitle{Optimal Quantized Correlation}

\author{M.\ D.\ Johnson, H.\ H.\ Chou, C.\ R.\ Gwinn}
\shortauthors{Johnson, Chou, \& Gwinn}
\affil{Department of Physics, University of California, Santa Barbara, California 93106, USA}
\email{ michaeltdh@physics.ucsb.edu,cgwinn@physics.ucsb.edu} 

\begin{abstract}
Using a maximum-likelihood criterion, we derive optimal correlation strategies for signals with and without digitization. 
We assume that the signals are drawn from zero-mean Gaussian distributions, as is expected in radio-astronomical applications, and we present correlation estimators both with and without {\it a priori} knowledge of the signal variances. 
We demonstrate that traditional estimators of correlation, which rely on averaging products, exhibit large and paradoxical noise when the correlation is strong. 
However, we also show that these estimators are fully optimal in the limit of vanishing correlation. 
We calculate the bias and noise in each of these estimators and discuss their suitability for implementation in modern digital correlators.  
\end{abstract}


\section{Introduction}

Because astrophysical sources emit Gaussian noise, the information in astrophysical observations lies in signal covariances. These are familiar as power spectra and cross spectra \citep[see, for example,][]{tms}. 
The estimation of these covariances is subject to bias and noise, and techniques to minimize both are therefore fundamental to radio astronomy. 

This estimation is complicated by the typically aggressive quantization of the received signal. Even for next-generation phased arrays, such as the Square-Kilometer Array, the cost of signal transmission necessitates low-bit quantization \citep{SKA_Dewdney}. This procedure distorts the spectrum but preserves much of the underlying statistical information. In fact, several authors have noted the ability of quantization to \emph{improve} estimates of correlation $\rho \in [-1,1]$, especially for strong correlation $|\rho| \rightarrow 1$ \citep{gwinn04}. Indeed, \citet{Cole_68} found that the standard two-level correlation scheme has lower noise than four- and six-level schemes in this limit, and that all of these estimates have lower noise than the correlation estimates for unquantized signals. 
We demonstrate that this paradoxical behavior arises from two causes: comparisons with unquantized correlation estimates are incomplete, and typical quantization schemes are not optimal. 

To amend these deficiencies, we present a correlation estimator for unquantized data that is appropriately suited to define a quantization efficiency, and we derive optimal correlation estimators for quantized signals via a maximum-likelihood criterion. With the recent advent of digital correlators in radio astronomy, such as DiFX \citep{DiFX}, implementing these techniques is straightforward.

\subsection{Terminology and Notation}

In spite of the many treatments of quantized correlation, no standard terminology has been adopted, so we first outline some basic assumptions and definitions. Throughout this work, we use $\{x_i,y_i\}$ to designate sets of pairs independently drawn from a zero-mean bivariate Gaussian distribution. For simplicity, we will assume that the standard deviations $\sigma_{\rm x}$ and $\sigma_{\rm y}$ are unity. 

%
We denote ensemble averages by unsubscripted angular brackets $\langle \ldots \rangle$. We will make use of the \emph{correlation} $\rho \equiv \langle x y \rangle /(\sigma_{\rm x} \sigma_{\rm y})$ and the \emph{covariance} $\langle x y \rangle$.

We denote finite averages, over a sample of $N$ points, by subscripted angular brackets $\langle \ldots \rangle_N$. For example, we frequently use the \emph{sample covariance} $r_{\infty} \equiv \langle x y \rangle_N = N^{-1} \sum_{i=1}^N x_i y_i$. We also use this terminology to refer to an average product after quantization. Because most applications of correlation in radio astronomy involve many samples $N$, we focus on the large--$N$ regime. 

Our work focuses on estimators of $\rho$, given a set of $N$ samples $\{ x_i, y_i \}$, possibly after quantization. We use the variable $r$, with subscripted identifiers, to indicate such estimators. Finally, we generically use $P(\ldots)$ to denote a probability density function (PDF) with respect to the given variables and parameters.

\subsection{Relation to Previous Work}

Previous analyses of quantized correlation have assumed that the correlation should be estimated via a form of sample covariance for the quantized signals; they have then optimized the performance of the correlation by choosing an appropriate quantization scheme. Furthermore, these efforts generally focus on the small correlation regime: $|\rho| \ll 1$. 

For example, \citet{ja98} provide an approximate prescription for correcting the bias from quantization in sample covariance. However, this prescription still suffers from severely sub-optimal performance when $\rho \neq 0$, in terms of the noise. 

In contrast, we provide a new mechanism for estimating correlation and demonstrate that it provides the lowest RMS error of \emph{any} post-quantization correlation strategy for a large number of samples. We also demonstrate this this strategy is equivalent to traditional approaches as $\rho \rightarrow 0$, and we give a rigorous justification for the optimal weights that are typically quoted.

\subsection{Outline of Paper}

In \S\ref{sec::Mathematical}, we briefly review the basic mathematical framework of parameter estimation theory, and we define the sense in which a particular strategy can be ``optimal.'' 
Then, in \S\ref{sec::Unquantized}, we consider the case of unquantized signals and present the corresponding optimal estimators for correlation. 
Next, in \S\ref{sec::Traditional_Quantized}, we summarize the details of the quantization procedure, outline the traditional correlation estimators via sample covariance, and derive the maximum-likelihood estimate of correlation for quantized signals. 
In \S\ref{sec::Examples}, we give specific examples for common quantization schemes, and compare the performance of the maximum-likelihood estimate to that of traditional estimates. 
Then, in \S\ref{sec::small_corr}, we demonstrate the critical property that traditional correlation schemes are optimal for small $|\rho|$. 
Finally, in \S\ref{sec::Summary}, we summarize our findings and discuss the possibilities for implementation.

\section{Mathematical Background}
\label{sec::Mathematical}

We begin by reviewing some essential concepts and terminology in parameter estimation theory. For a comprehensive discussion of these ideas with a rigorous description of the assumptions and constraints, see \citet{Kendall_Stuart_v2} or \citet{Lehmann_Casella}.

\subsection{Optimal Estimators and Maximum Likelihood}
\label{sec::Estimator_Background}

We first ascribe a precise meaning to the term ``optimal'' estimator. For this purpose, we must consider both the bias and noise in an estimator. 
We seek estimates of correlation that converge to the exact correlation as $N \rightarrow \infty$; such estimates are said to be \emph{consistent}. We refer to a consistent estimator with the minimum noise (i.e.\ the minimum mean squared error) as the optimal estimator.

If the parameters to be estimated correspond to a known class of distributions, then a particularly simple estimator can be defined. Namely, consider a set of observations $\textbf{x}$ drawn from a distribution that is specified by a set of parameters $\boldsymbol{\theta}_0$. One parameter estimation strategy determines the parameters which maximize the likelihood function $\mathcal{L}(\boldsymbol{\theta}|\textbf{x})$, defined as the probability of sampling $\textbf{x}$ given the distribution specified by $\boldsymbol{\theta}$. If $\mathcal{L}$ has a unique maximum at some $\hat{\boldsymbol{\theta}}_{\rm ML}$, then this point is defined to be the \emph{maximum-likelihood estimator} (MLE) of $\boldsymbol{\theta}_0$ for the sampled points $\textbf{x}$. 

Often, the sample data $\textbf{x}$ can be greatly reduced to some simplified statistic $\textbf{T}(\textbf{x})$, which carries all the information related to the parameters $\boldsymbol{\theta}_0$. In this case, $\textbf{T}(\textbf{x})$ is said to be a \emph{sufficient} statistic for $\boldsymbol{\theta}_0$. For example, if samples are drawn from a normal distribution with known variance but unknown mean, then the sample mean is a sufficient statistic for the mean. The \emph{factorization criterion} states that a necessary and sufficient condition for $\textbf{T}(\textbf{x})$ to be sufficient for a family of distributions parametrized by $\boldsymbol{\theta}_0$ is that there exist non-negative functions $g$ and $h$ such that $P(\textbf{x};\boldsymbol{\theta}_0) = g[\textbf{T}(\textbf{x});\boldsymbol{\theta}_0] h(\textbf{x})$.

Under weak regularity conditions, the likelihood function also determines the minimum noise that \emph{any} unbiased estimator can achieve. This minimum, the Cram\'{e}r-Rao bound (CRB), can be expressed in terms of derivatives of $\mathcal{L}$. For example, the minimum variance of any unbiased estimator $\hat{\theta}$ of a single parameter $\theta_0$ is the inverse of the \emph{Fisher information}, and can be written
\begin{align}
\left \langle \delta \hat{\theta}^2 \right \rangle \geq\left \langle \left( \left. \frac{\partial \ln \mathcal{L}(\textbf{x};\theta)}{\partial \theta} \right \rfloor_{\theta=\theta_0} \right)^2 \right \rangle^{-1}\! \equiv \delta \hat{\theta}^2_{\rm CR}.
\end{align}
Here, $\langle \ldots \rangle$ denotes an ensemble average over sets of measurements $\textbf{x}$. 
An unbiased estimator with noise that matches the CRB is said to be \emph{efficient}.

Under general conditions, the MLE is both consistent and asymptotically (as $N {\rightarrow} \infty$) efficient. In the present work, we present the MLE of correlation for both unquantized and quantized signals, and we compare these correlation strategies with traditional schemes.

\subsection{Distribution of Correlated Gaussian Variables}
Astrophysical observations measure zero-mean, Gaussian noise. Under rather broad assumptions, pairs of such samples $\{x, y\}$ are drawn from a bivariate Gaussian distribution. In addition to the respective standard deviations, $\sigma_{\rm x} \equiv \sqrt{\langle x^2 \rangle}$ and $\sigma_{\rm y} \equiv \sqrt{\langle y^2 \rangle}$, this distribution depends on the correlation $\rho \equiv \langle x y \rangle/(\sigma_{\rm x} \sigma_{\rm y}) \in [-1,1]$. Because our present emphasis is correlation, we assume that $\sigma_{\rm x} = \sigma_{\rm y} = 1$, in which case the PDF is given by
\begin{align}
\label{eq::bivariate}
P(x,y;\rho) = \frac{1}{2\pi\sqrt{1-\rho^2}} \exp\left[ -\frac{x^2 + y^2 - 2\rho x y  }{2\left(1-\rho^2\right)} \right].
\end{align}
For small $|\rho|$, this distribution takes the following approximate form:
\begin{align}
\label{eq::bivariate_approx}
P(x,y;\rho) \approx \frac{1}{2\pi}\left( 1 + \rho x y \right) e^{-\frac{1}{2}\left( x^2 + y^2 \right)}.
\end{align}
See Chapter 8 of \citet{tms} (hereafter TMS) for some additional representations and discussion.

%

\section{Correlation Estimators for Unquantized Signals}
\label{sec::Unquantized}

We now analyze several correlation estimators for unquantized signals. These estimators serve two relevant purposes: they provide a point of reference to ascribe an efficiency to a quantization scheme, and they suggest closed-form strategies for correlation estimates of quantized signals that have a large number of bits.

\begin{figure}[t]
\includegraphics*[width=0.48\textwidth]{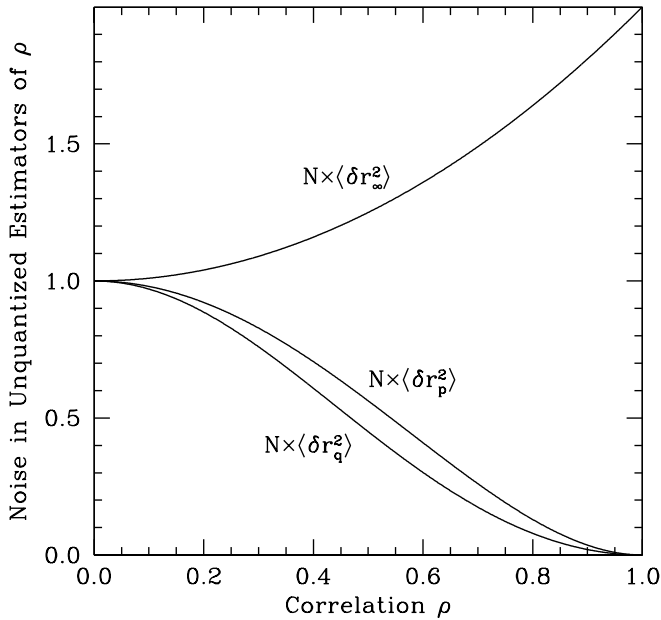}
\caption{
Asymptotic noise in estimators $r_{\infty}$, $r_{\rm p}$, and $r_{\rm q}$ of $\rho$ for unquantized signals. Because the noise is a symmetric function of $\rho$, only the positive values are shown.
}
\label{fig::corr_compare}
\end{figure}

First, in \S\ref{sec::r_inf}, we consider the estimate of correlation via sample covariance, denoted $r_{\infty}$. Next, in \S\ref{sec::r_p}, we present Pearson's estimate of correlation, $r_{\rm p}$, which serves as the optimal estimator when no information about the signal is known. Last, in \S\ref{sec::r_q}, we give details of the MLE of correlation when the signal variances are known, which we denote $r_{\rm q}$. Figure \ref{fig::corr_compare} compares the asymptotic noise in these three estimates, as given in the following sections.

\subsection{Correlation via Sample Covariance: $r_{\infty}$}
\label{sec::r_inf}
The simplest estimate of correlation follows from the relationship between correlation and covariance. Namely, suppose that the means $\{\mu_{\rm x},\mu_{\rm y} \}$ and standard deviations $\{\sigma_{\rm x},\sigma_{\rm y}\}$ of the signals $x_i$ and $y_i$ are known. 
In this case, the signals may be standardized to have zero mean and unit variance. Their covariance is then equal to their correlation: $\langle x y \rangle = \rho$. This correspondence immediately suggests a simple estimator for the correlation: $r_{\infty} \equiv \langle x y \rangle_N = N^{-1} \sum_{i=1}^N x_i y_i$. 

This estimator is unbiased and consistent but has large variance: $N \langle \delta r_{\infty}^2 \rangle \equiv N \langle (r-\rho)^2 \rangle = \left(1+\rho^2\right)$ \citepalias[Eq.\ 8.13]{tms}. This equation highlights a peculiar feature: $r_{\infty}$ is the noisiest when the correlation is \emph{strongest}. 

\subsection{Optimal Correlation: $r_{\rm p}$ }
\label{sec::r_p}

Many researchers have studied improved estimators of correlation (see \citet{h53} and \citet{a96} for interesting perspectives). The most common estimator is known as ``Pearson's r'' and is given by
\begin{align}
r_{\rm p} \equiv \frac{\left \langle \left( x - \left \langle x \right \rangle_N \right)\left( y - \left \langle y \right \rangle_N \right) \right \rangle_N}{\sqrt{\left \langle \left( x - \left \langle x \right \rangle_N \right)^2\right \rangle_N \left \langle \left( y - \left \langle y \right \rangle_N \right)^2 \right \rangle_N}}.
\end{align}
In addition to being unbiased and consistent, $r_{\rm p}$ is asymptotically efficient. The asymptotic noise can be derived using the Fisher transformation \citep{f1915,f1921}: $\lim_{N \rightarrow \infty} N \left \langle \delta r_{\rm p}^2 \right \rangle = \left( 1 - \rho^2 \right)^2$, which is indeed the CRB.

Note that substituting the exact means and variances into $r_{\mathrm{p}}$ returns the original estimate $r_{\infty}$ and, remarkably, \emph{decreases} the quality of the estimate. As a simple example, three randomly generated samples with correlation $\rho = 0.999$ are $\{x_i\} = \{0.998,1.712,-0.992\},\{y_i\} = \{1.01,2.01,-0.980\}$. In this case, we obtain estimates $r_{\infty} = 1.81$, $r_{\rm p} = 0.997$. Indeed, for perfect correlation, $\delta r_{\rm p}^2 = 0$, whereas $\delta r_{\infty}^2 = 2/N$. Pearson's estimate accounts for the sample variance, which contributes much of the noise in $r_{\infty}$. However, for small correlation, $\rho \rightarrow 0$, the noise in $r_{\infty}$ and $r_{\rm p}$ is identical (we further discuss this feature in \S\ref{sec::small_corr}).

Hence, when the correlation is large, simply averaging products poorly approximates the correlation relative to other schemes. Even if the exact variance is known, the sample variance must still be incorporated to optimally estimate the correlation. 

\subsection{Optimal Correlation with Known Signal Variance: $r_{\rm q}$ }
\label{sec::r_q}

Nevertheless, an exact knowledge of the variance can be used to effectively improve the estimate of correlation. In fact, this knowledge is generally assumed in radio astronomy. For example, automatic gain control usually sets the variances $\langle x^2\rangle= \langle y^2 \rangle = 1$, and quantization schemes use the ``known'' variance to determine the appropriate level settings. Errors in the signal estimate are then a source of both bias and noise, so for non-stationary signals such as pulsars, the quantization weights must be dynamically adjusted \citep[see][]{ja98}. For a more complete discussion of quantization noise, see \citet{gwinn04} and \citet{gwinn06}.

More generally, whenever the timescale of variation of $\rho$ is shorter than that of variation in the standard deviation $\sigma$ of $x$ and $y$, there will be improved measures of correlation. 

For example, if the standard deviations of $x$ and $y$ are known to be unity, then the MLE of correlation, denoted $r_{\rm q}$, is determined by the real solution of
\begin{align}
r_{\infty} \left(1+r_{\rm q}^2\right) - r_{\rm q} \left( \langle x^2 \rangle_N + \langle y^2 \rangle_N - \left(1-r_{\rm q}^2\right)\right) = 0.
\end{align}
In \S\ref{seq::r_q_deriv}, we derive this result, give an approximate form for $r_{\rm q}$, and demonstrate that the noise in this estimate achieves the CRB as $N\rightarrow \infty$, as expected for an MLE: $\lim_{N \rightarrow \infty} N \left \langle \delta r_{\rm q}^2 \right \rangle = \left( 1 - \rho^2 \right)^2\! / \left(1 + \rho^2\right)$. The advantage of $r_{\rm q}$ relative to $r_{\rm p}$ increases with $|\rho|$, and gives a factor of two improvement in the estimator variance at high correlation. Moreover, the bias in $r_{\rm q}$ is $o\left(N^{-1}\right)$.

\section{Correlation Estimators for Quantized Signals}
\label{sec::Traditional_Quantized}

In practice, data are digitized, which involves quantization according to a prescribed scheme. The merit of the quantization, reduction of data volume, must be carefully weighed against its drawback, degraded signal information. 

We first review the details of quantization and the traditional estimators of correlation, which rely on the sample covariance after quantization. We then derive the MLEs of correlation for arbitrary quantization schemes and give expressions for the noise in these estimators.

\subsection{The Quantization Transfer Function}
The process of quantization maps each element in a time-series $x_i \in \mathbb{R}$ to a set of $L = 2^b$ discrete values: $x_i \mapsto \hat{x}_{L,i}$, where $b$ is the number of bits in the quantization scheme. This transfer function involves $L - 1$ thresholds, which partition $\mathbb{R}$ into $L$ subsets, and $L$ respective weights for these subsets.

\subsection{Quantized Correlation via Sample Covariance}

The traditional correlation estimator for quantized signals matches the form of the continuous covariance estimator, $r_{\infty}$, to the quantized signals $\hat{r}_L \equiv \langle \hat{x}_L \hat{y}_L \rangle_N$ \citep{vvm66,Cole_68,Cooper70,Hagen_Farley_73}. In some cases, this result is then appropriately transformed to account for bias. This correlation strategy is optimized through the particular thresholds and weights that determine the transfer function of the quantization.

\subsection{The MLE of Correlation for Quantized Signals}
\label{sec::MLEQ_D}

We now derive the MLE of correlation $r_{L,\mathrm{ML}}$ for quantized signals, which is both consistent and asymptotically efficient. In particular, the likelihood function $\mathcal{L}$ for a set of $N$ independent and identically distributed (i.i.d.) pairs of samples $\{ \hat{x}_{L,i},\hat{y}_{L,i} \}$ drawn from a bivariate normal distribution and then quantized in a scheme with $L$ levels is
\begin{align}
\label{eq::Quantized_L}
\mathcal{L}(\rho,\sigma | \{ \hat{x}_{L,i}, \hat{y}_{L,i} \}) &= \prod_{i=1}^N P(\hat{x}_{L,i},\hat{y}_{L,i};\rho,\sigma)\\
\Rightarrow \nonumber \ln \mathcal{L}(\rho,\sigma | \{ \hat{x}_{L,i}, \hat{y}_{L,i} \}) &= \sum_{\ell} \mathcal{N}_{\ell} \ln \mathcal{P}_{\ell}(\rho,\sigma).
\end{align}
In this expression, $\ell$ runs over the $L^2$ possible quantized pairs $\{ \hat{x}_L, \hat{y}_L \}$; $\mathcal{N}_{\ell}$ is the total number of samples in each such category; and $\mathcal{P}_{\ell}(\rho,\sigma)$ corresponds to the probability of a sampled pair $\{ \hat{x}_L,\hat{y}_L \}$ falling in that category. 

To determine the MLE, this log-likelihood must be maximized with respect to $\rho$, if $\sigma$ is assumed to be known, or with respect to $\rho$ and $\sigma$, if $\sigma$ is unknown. Although we have assumed symmetry $\sigma_{\rm x} = \sigma_{\rm y}$, the generalization is straightforward.

The MLE thus requires an evaluation of each probability $\mathcal{P}_{\ell}$:
\begin{align}
\label{eq::MLE_P}
\mathcal{P}_{\ell} = S_{\ell} \int_{R_{\ell}} dx dy\, P(x,y;\rho,\sigma).
\end{align}
In this expression, $P(x,y;\rho,\sigma)$ is given by Eq.\ \ref{eq::bivariate}, $R_{\ell} \subset \mathbb{R}^2$ corresponds to the set of unquantized values that map to each quantized state, and $S_{\ell}\in \mathbb{Z}$ is an optional symmetry factor, to account for the symmetry under inversion, $P(\hat{x}_{L,i},\hat{y}_{L,i}) = P(-\hat{x}_{L,i},-\hat{y}_{L,i})$, and transposition, $P(\hat{x}_{L,i},\hat{y}_{L,i}) = P(\hat{y}_{L,i},\hat{x}_{L,i})$. In a few instances, such as the quadrant integrals that arise in one-bit correlation, Eq.\ \ref{eq::MLE_P} has a simple, closed-form representation. More generally, it can be reduced to a one-dimensional integral of an error function.

Thus, in most cases, the MLE requires minimization over a function that involves one-dimensional numerical integration. However, many strategies can simplify this estimation. For example, if both the number of samples $N$ and quantization bits $b$ are small, then all required solutions can be tabulated. 
After including the symmetry reductions, the number of distinct correlation possibilities is $N_{\ell} = 2^{b-1}\left( 1 + 2^{b-1} \right)$. 
The total number $M$ of partitions of $N$ samples into these categories is then $M = \binom{N + N_{\ell} - 1}{N_{\ell}-1} \sim N^{N_{\ell}-1}/(N_{\ell}-1)!$. If $M$ is prohibitively large, then the $N$ samples can first be partitioned and then the respective correlation estimates averaged to obtain an approximation of the MLE.

\subsection{Noise in the MLE and the Cram\'{e}r-Rao Lower-Bound}
\label{sec::Quantized_CRB}

As we have already mentioned, the CRB determines the minimum variance that any unbiased estimator of $\rho$ can achieve. In terms of the likelihood function of \S\ref{sec::MLEQ_D}, the elements of the $2{\times}2$ Fisher information matrix for $\{\rho,\sigma\}$ can be written
\begin{align}
\label{eq::FisherI}
\mathcal{I}_{1,1} &\equiv \left \langle \left( \frac{\partial}{\partial \rho} \sum_{\ell} \mathcal{N}_{\ell} \ln \mathcal{P}_{\ell} \right)^2 \right \rangle = N \sum_{\ell} \frac{ \left(\frac{\partial \mathcal{P}_{\ell}}{\partial \rho} \right)^2 }{\mathcal{P}_{\ell}},\\
\nonumber \mathcal{I}_{1,2} &= N \sum_{\ell} \frac{ \left(\frac{\partial \mathcal{P}_{\ell}}{\partial \rho} \right)  \left(\frac{\partial \mathcal{P}_{\ell}}{\partial \sigma} \right) }{\mathcal{P}_{\ell}},\\
\nonumber \mathcal{I}_{2,2} &= N \sum_{\ell} \frac{ \left(\frac{\partial \mathcal{P}_{\ell}}{\partial \sigma} \right)^2 }{\mathcal{P}_{\ell}}.
\end{align}
If $\sigma$ is known, then the minimum variance of an unbiased estimator of $\rho$ is $\delta r_{L,\mathrm{CR}}^2 = \mathcal{I}_{1,1}^{-1}$; if $\sigma$ is unknown, then the minimum variance is $\delta r_{L,\mathrm{CR}}^2 = \mathcal{I}_{2,2}/\left(\mathcal{I}_{1,1} \mathcal{I}_{2,2} - \mathcal{I}_{1,2}^2 \right)$. The MLE is asymptotically efficient, so $\langle \delta r_{L,\mathrm{ML}}^2 \rangle \rightarrow \delta r_{L,\mathrm{CR}}^2$ as $N\rightarrow \infty$.

\section{Examples}
\label{sec::Examples}

\subsection{One-bit Quantization}
In the standard one-bit, or two-level, quantization scheme, each sample is reduced to one ``sign'' bit: $x \mapsto \hat{x}_2 \equiv \mathrm{sign}(x)$. 
Because the sample error for the signal variance incurs the bulk of the noise in $r_{\infty}$, quantization actually \emph{improves} upon the estimate of $r_{\infty}$ in some cases. 

Explicitly, we have $r_2 \equiv \left \langle \hat{x} \hat{y} \right \rangle_N$. However, this estimate is biased: $\left \langle r_2 \right \rangle = 2\pi^{-1} \sin^{-1} \rho$. The standard Van Vleck clipping correction, denoted $r_{2,\mathrm{V}}$, improves the bias to $\mathcal{O}(1/N)$ by simply inverting this relationship \citep{vvm66}:
\begin{align}
r_{2,\mathrm{V}} \equiv \sin \left( \frac{\pi}{2} r_2 \right).
\end{align}

In fact, $r_{2,\mathrm{V}}$ gives precisely the MLE. To see this, note that the quantized products, $\hat{x} \hat{y}$, have probability $P(\pm 1) = \frac{1}{2} \pm \frac{1}{\pi} \arcsin{\rho}$. Minimizing the log-likelihood (Eq.\ \ref{eq::Quantized_L}) with respect to $\rho$ gives that $r_{2,\mathrm{ML}} = r_{2,\mathrm{V}}$. 

Because $r_{2,\mathrm{V}}$ is the MLE, the noise for large $N$ is given by the CRB. Substituting $P(\pm 1)$ into Eq.\ \ref{eq::FisherI} gives
\begin{align}
\label{eq::r2_CRB}
N \delta r_{2,\mathrm{CR}}^2 = \left[\left(\frac{\pi}{2}\right)^2 - ( \arcsin{\rho} )^2 \right] \left(1 - \rho^2\right).
\end{align}
We can easily verify that the noise in $r_{2,\mathrm{V}}$ actually achieves this lower bound. Namely, the correlation estimate $r_2$ is a one-dimensional random walk with $N$ steps of length $\pm 1/N$, distributed according to $P(\pm 1)$. For large $N$, the central limit theorem gives that $r_2$ follows a Gaussian distribution with mean $2\pi^{-1} \arcsin{\rho}$ and variance $N^{-1}\left[1-\left(2\pi^{-1}\arcsin{\rho}\right)^2 \right]$. In this limit, we obtain
\begin{align}
\label{eq::r2V_noise}
\nonumber \left \langle r_{2,{\rm V}}^2 \right \rangle = \frac{1}{2} \left\{ 1 - \left(1 - 2\rho^2 \right) \exp \left[\frac{4 (\arcsin{\rho})^2 - \pi^2}{2N} \right] \right\}\\
\Rightarrow N \left \langle \delta r_{2,{\rm V}}^2 \right \rangle \approx \left[\left(\frac{\pi}{2}\right)^2 - ( \arcsin{\rho} )^2 \right] \left(1 - \rho^2\right),
\end{align}
which is identical to the CRB.

The most striking improvement of $r_{2,\mathrm{V}}$ relative to $r_{\infty}$ occurs as $\rho \rightarrow \pm 1$; in this limit, the one-bit correlation has no noise, while $\langle \delta r_{\infty}^2 \rangle = 2/N$.

\subsection{Two-bit Quantization}

Perhaps the most common quantization strategy replaces each sample by a pair of bits for sign and magnitude. The (non-zero) thresholds $\pm v_0$ are fixed at some level relative to the estimated RMS signal voltage $\sigma$ in a way that minimizes the expected RMS noise in the subsequent correlation estimates. The resulting four levels are then assigned weights $\hat{x}_2\in \{ \pm 1,\pm n\}$, where $n$ is also chosen to minimize the noise. In terms of the mean quantized product $r_4 \equiv \langle \hat{x}_4 \hat{y}_4 \rangle_N$, one obtains the correlation estimate \citepalias[Eq.\ 8.43]{tms}
\begin{align}
r_{4,\mathrm{v}} = \frac{r_4}{\Phi + n^2(1-\Phi)},
\qquad \Phi \equiv \mathrm{erf}\left( \frac{v_0}{\sigma \sqrt{2}} \right).
\end{align}
This estimate of correlation, which already assumes exact knowledge of $\sigma$, retains a significant (${\sim} 10\%$) bias to high $|\rho|$ (see Figure 1 of \citet{ja98}). If $|\rho| \lsim 0.8$, for instance, then the appropriate correction is simply a constant scaling factor \citepalias[Eq.\ 8.45]{tms}:
\begin{align}
r_{4,\mathrm{V}} \approx \left\{ \frac{\pi \left[\Phi + n^2 (1-\Phi) \right]}{2\left[ \left(n-1 \right) E + 1 \right]^2} \right\}  r_{4,\mathrm{v}},
\qquad E \equiv e^{-\frac{1}{2} \left( \frac{v_0}{\sigma} \right)^2}.
\end{align}
For additional details and a complete formulation to remove the bias, see \citet{gwinn04}. Here, we use the ``V'' subscript to draw analogy with the Van Vleck correction for one-bit correlation. Namely, this estimate calculates the sample covariance after quantization and then performs a bias correction according to the estimated correlation. The remaining bias is $\mathcal{O}(1/N)$.

Researchers then optimize this two-bit correlation scheme by a particular choice of thresholds and weights: $v_0 = 0.9815$, $n=3.3359$. However, unlike one-bit correlation, the bias-corrected quantized product $r_{4,\mathrm{V}}$ is \emph{not} the optimal estimator of correlation for quantized data. In particular, $r_{4,\mathrm{V}}$ even reflects the disturbing feature of the continuous estimate $r_{\infty}$ that the noise tends to increase with $|\rho|$. Hence, high correlations present the paradoxical situation in which traditional estimates $\{ r_{\infty} ,r_{4,\mathrm{V}},r_{2,{\mathrm{V}}}\}$ perform \emph{better} as the number of bits is \emph{reduced}.

\begin{figure}[t]
\includegraphics*[width=0.48\textwidth]{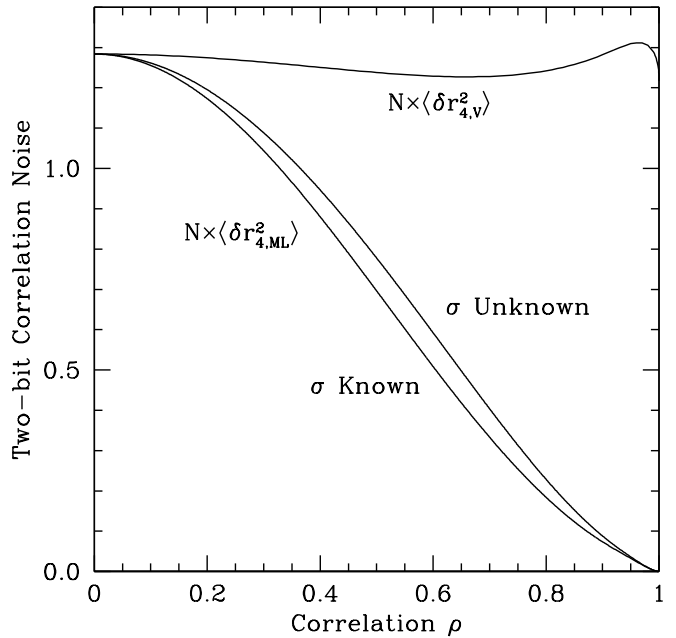}
\caption{
Noise in estimates of correlation for signals quantized with two bits. The chosen levels ($v_0 = 0.9815$) and weights ($n=3.3359$) are optimal as $|\rho| \rightarrow 0$. The upper curve gives the noise in the traditional estimator via sample covariance, as derived in \citet{gwinn04}, whereas the lower curves give the noise in the MLEs with and without knowledge of $\sigma$.
}
\label{fig::twobit_noise}
\end{figure}

This troubling evolution merely reflects the incompleteness of these correlation estimates. Figure \ref{fig::twobit_noise} compares the noise in $r_{4,\mathrm{V}}$ to the noise in the MLE, both when $\sigma$ is known and unknown. Each MLE has negligible bias and noise that reflects the behavior seen in the corresponding unquantized MLE, $r_{\rm p}$ or $r_{\rm q}$. The noise is always lower than that of $r_{2,\mathrm{V}}$ and approaches zero as $|\rho| \rightarrow 1$. We therefore resolve the puzzling evolution of correlation noise after quantization. 

The only remaining barrier is the computational difficulty of implementation. However, for small values of $N$, the maximum-likelihood solutions may be tabulated prior to calculation; the required number of tabulated values is $M = \binom{N + 5}{5} \sim N^{5}/5!$ (see \S\ref{sec::MLEQ_D}). Alternatively, one can first partition the $N$ samples, then calculate the MLE of correlation for each subset via tabulation, and finally average the results. We defer a comprehensive treatment of these implementation strategies to a future work.

\subsection{Many-bit Quantization}

\begin{figure}[t]
\centering 
\includegraphics*[width=0.48\textwidth]{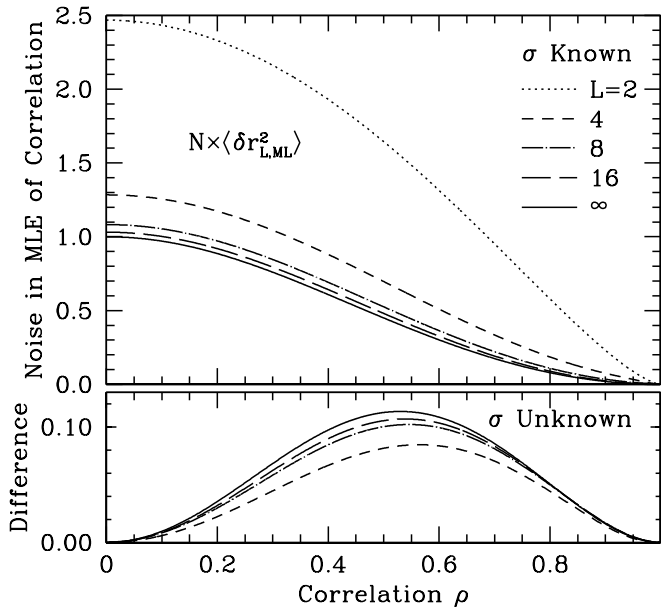}
\caption{
Noise in the MLE of correlation for quantized signals as a function of correlation for various quantization levels. The upper panel shows the noise when $\sigma$ is known, and the lower panel shows the additional noise when $\sigma$ is unknown. For simplicity, we set the $(L-1)$ quantization thresholds in multiples of $\pm 4/L$. Observe that the reduction in noise provided by knowledge of $\sigma$ becomes more pronounced as the number of levels is \emph{increased}. For example, knowledge of $\sigma$ provides no improvement for one-bit correlation. 
}
\label{fig::quantization_noise}
\end{figure}

Modern instrumentation now permits the storage of baseband data with many-bit quantization schemes. 
In this case, the noise in the MLE of correlation rapidly approaches that in the corresponding unquantized limit, $r_{\rm p}$ or $r_{\rm q}$ (see Figure \ref{fig::quantization_noise}). In such cases, these estimators for unquantized signals provide excellent approximations of the quantized MLEs, and the primary concerns are the influence of RFI and instrumental limitations \citepalias{tms}. Furthermore, although low-bit quantization schemes are quite robust to impulsive RFI, estimates such as $r_{\rm p}$ are not, so alternative quantization schemes that are robust at the expense of increased noise may be preferred \citep{fridman_robust}.

We now consider the incurred bias when approximating the quantized MLE by $r_{\rm p}$. Specifically, consider a high-bit scheme with $L$ levels, thresholds in multiples of $\pm v_0$, and quantization weights $\hat{x}$ that are the average values of their respective preimages. We denote the corresponding estimator $r_{L,\mathrm{p}}$. Then, if the highest thresholds extend far into the tail of the distribution, the bias after quantization is approximately
\begin{align}
\delta r_{L,\mathrm{p}} \approx -\frac{1}{12}\left(\frac{v_0}{\sigma}\right)^2 \rho.
\end{align}
For more general expressions, which include the effects of the finite outer thresholds, consult the discussion in \S8.3 of \citetalias{tms}. While correcting the bias is straightforward, even for a low number of bits, this strategy is ineffective for low-bit schemes because $r_{L,\mathrm{p}}$ is not a sufficient statistic for $\rho$.

\section{Reductions for Small Correlation}
\label{sec::small_corr}

Although the MLE of correlation decreases the noise for large $|\rho|$, it exhibits identical noise to traditional estimators at small $|\rho|$. Furthermore, in this limit, knowledge of $\sigma$ does not reduce the noise. These features both arise from the form of the bivariate Gaussian PDF in this limit. 

For example, consider the estimation of correlation for unquantized signals when $\sigma_{\rm x} = \sigma_{\rm y} = 1$ is known. Then, if $|\rho|\ll 1$, the joint PDF of $N$ independently-drawn pairs of correlated random variables $\{x_i,y_i\}$ is (see Eq.\ \ref{eq::bivariate_approx})
\begin{align}
P(\{x_i,y_i\};\rho) \approx \frac{1}{(2\pi)^N}\left( 1 + N \rho \langle x y \rangle_N \right) e^{-\frac{N}{2}\left( \langle x^2 \rangle_N + \langle y^2 \rangle_N \right)}.
\end{align}
Thus, from the factorization criterion, $r_{\infty} \equiv \langle x y \rangle_N$ is a sufficient statistic for $\rho$, and so we expect the asymptotic noise in $r_{\infty}$ to match that of $r_{\rm q}$ as $\rho\rightarrow 0$. 

Likewise, consider the joint distribution of the samples after quantization into $L$ weighted levels. In this case, we require the set of quantized probabilities
\begin{align}
\mathcal{P}_{\ell} &\approx  \frac{1}{2\pi} \int_{R_\ell} dx dy\, \left(1 + \rho x y \right) e^{-\frac{1}{2}\left( x^2 + y^2 \right)} \equiv \alpha_{\ell} + \beta_{\ell} \rho.
\end{align}
The joint PDF of the quantized samples is then
\begin{align}
P(\{\hat{x}_{L,i},\hat{y}_{L,i}\};\rho) &= \prod_{\ell} \mathcal{P}_{\ell}^{\mathcal{N}_{\ell}}\\
\nonumber &\approx \left( \prod_{\ell} \alpha_{\ell}^{\mathcal{N}_{\ell}} \right) \left( 1 + \rho \sum_{\ell} \mathcal{N}_{\ell} \frac{\beta_{\ell}}{\alpha_{\ell}}  \right).
\end{align}
Hence, for small $|\rho|$, the factorization criterion gives that $\left\langle w(\hat{x},\hat{y}) \right\rangle$ is a sufficient statistic for $\rho$, if the weight function is determined by
\begin{align}
\label{eq::quant_fact}
w(\hat{x},\hat{y}) &= \frac{ \int_{R_\ell} dx dy\, x y e^{-\frac{1}{2}\left( x^2 + y^2 \right)} }{ \int_{R_\ell} dx dy\, e^{-\frac{1}{2}\left( x^2 + y^2 \right)} }\\
\nonumber &= \left[ \frac{ \int_{R_{\ell,x}} dx \, x e^{-\frac{x^2}{2}} }{ \int_{R_{\ell,x}} dx\, e^{-\frac{x^2}{2}} } \right] \left[ \frac{ \int_{R_{\ell,y}} dy \, y e^{-\frac{y^2}{2}} }{ \int_{R_{\ell,y}} dy\, e^{-\frac{y^2}{2}} } \right],
\end{align}
where $R_{\ell,x}\subset \mathbb{R}$ defines the range of values spanned by each quantized level. 

Moreover, the final factorization in Eq.\ \ref{eq::quant_fact} demonstrates that, by assigning an appropriate weight to each quantization level: $w(\hat{x},\hat{y}) = \hat{x} \hat{y}$, the sample covariance is a sufficient statistic for $\rho$ and will achieve optimal noise performance as $|\rho| \rightarrow 0$. 

The asymptotic noise in this limit is then the CRB:
\begin{align}
\left. \delta r_{L,\mathrm{CR}}^2\right \rfloor_{\rho=0} &= \frac{2\pi}{N} \left\{ \sum_{\ell} \frac {\left[ \int_{R_\ell} dx dy \, x y  e^{-\frac{1}{2}\left( x^2 + y^2 \right)}\right]^2 }{ \int_{R_\ell} dx dy\,  e^{-\frac{1}{2}\left( x^2 + y^2 \right)} } \right \}^{-1}\!.
\end{align}
Minimizing this equation yields the optimal thresholds. Then, Eq.\ \ref{eq::quant_fact} immediately determines the optimal weights. Observe that these weights are slightly different than those of some previous works, such as \citet{ja98}, but match the ratios of traditional quantization schemes, such as $n=3.336$ when $v_0 = 0.982$ for two-bit correlation, for instance.

Finally, $\mathcal{I}_{1,2} \rightarrow 0$ as $\rho \rightarrow 0$. This result follows easily by substituting $\mathcal{P}_{\ell}$ and its derivatives into Eq.\ \ref{eq::FisherI}. Hence, the CRB is unchanged by knowledge of $\sigma$ in this limit.

\section{Summary}
\label{sec::Summary}

We have explored the paradoxical scaling of noise in traditional estimates of correlation for quantized signals. In particular, we have shown that the decrease in noise that quantization affords is a result of an incomplete comparison with unquantized correlation schemes and of sub-optimal correlation strategies for quantized signals. 

We have derived the MLE of correlation, both with and without knowledge of the signal variance and quantization, and we have compared these estimates to traditional schemes. The MLE has negligible bias, lower noise, and is asymptotically efficient: for a large number of samples, no other unbiased scheme will achieve lower noise. We have also derived simple expressions for this asymptotic noise (the CRB). While the MLE gives the familiar Van-Vleck corrected sample covariance for one-bit quantization, it differs significantly from current correlation strategies for all other cases.

Nevertheless, traditional correlation schemes are fully optimized in the limit $\rho \rightarrow 0$. Namely, for suitably chosen weights, the sample covariance $\hat{r}_L$ is a sufficient statistic for the correlation $\rho$, in this limit. 

Future detectors, such as the Square-Kilometer Array, that will achieve high signal-to-noise while being limited to a small number of quantization bits, can benefit from these novel correlation strategies to reduce both the distortion and noise introduced by quantization.

\acknowledgments
We thank the U.S. National Science Foundation for financial support for this work (AST-1008865). The work of H.\ C.\ was supported by a UCSB SURF award.

\appendix

\section{MLE for Unquantized Signals with Known Variance}
\label{seq::r_q_deriv}

We now summarize the main features of the MLE of correlation for samples $\{x_i, y_i\}$ drawn from a bivariate Gaussian distribution with known means and variances. See \citet{Kendall_Stuart_v2} for additional details. For simplicity, we assume that the means are zero and the variances are unity. We also assume that each pair is drawn independently. The likelihood function is then 
\begin{align}
\mathcal{L}(\rho | \{ x_i, y_i \}) \equiv \prod_{i=1}^N P(x_i,y_i;\rho,\sigma_{\rm x},\sigma_{\rm y})
= \frac{1}{2\pi \sqrt{1 - \rho^2} } \exp \left[ -\frac{1}{2\left(1 - \rho^2 \right)} \sum_{i=1}^N \left( x_i^2 + y_i^2 - 2\rho x_i y_i \right)  \right].
\end{align}
The condition for the likelihood function to be extremized is
\begin{align}
\label{newcorr}
r_{\infty} \left(1+\rho^2\right) - \rho \left( s_{\rm x}^2 + s_{\rm y}^2 - \left(1-\rho^2\right)\right) = 0,
\end{align}
where $s_{\rm x}^2 \equiv \langle x^2 \rangle_N$, $s_{\rm y}^2 \equiv \langle y^2 \rangle_N$, and $r_{\infty} \equiv \left \langle x y \right \rangle_N$. 
Hence, the triplet $\{r_{\infty},s_{\rm x},s_{\rm y}\}$ is sufficient for $\rho$. 
We will denote the appropriate solution to this cubic equation $r_{\rm q}$.

To obtain some intuition for this result, let $\epsilon \equiv s_{\rm x}^2 + s_{\rm y}^2 - 2$. Then $\langle \epsilon^2 \rangle = 4(1+\rho^2)/N$. The discriminant of the cubic is
\begin{align}
\Delta = -4 r_{\infty}^4 + \left( \epsilon^2 + 20 \epsilon - 8 \right) r_{\infty}^2 - 4 \left( 1 + \epsilon \right)^3.
\end{align}
If $\Delta<0$, then the cubic has a single real solution. As a rough rule of thumb, we can simply consider when all terms are negative. Since $\delta \epsilon \approx 2/\sqrt{N}$, we see that there is likely a unique real solution whenever $\epsilon < .39$, or $N \gtwid 25$.

Although finding this solution is both analytically and numerically straightforward, an approximation is both useful and enlightening:
\begin{align}
\label{eq::rq_expand}
r_{\rm q} = r_{\infty} \left[ 1 - \frac{1}{1+r_{\infty}^2} \epsilon + \frac{1-r_{\infty}^2}{\left(1+r_{\infty}^2\right)^3} \epsilon^2 + \mathcal{O}\left(\epsilon^3\right)\right].
\end{align}
This expansion immediately identifies the appropriate root of the cubic equation. 
Furthermore, we can determine the asymptotic noise for $r_{\rm q}$ by expanding Eq.\ \ref{eq::rq_expand} for large $N$:
\begin{align}
\left \langle \delta r_{\rm q}^2 \right \rangle = \left \langle \delta r_{\infty}^2 \right \rangle + \frac{\rho^2}{\left(1 + \rho^2 \right)^2} \left \langle \epsilon^2 \right \rangle - \frac{2 \rho}{1+\rho^2} \left \langle \delta r_{\infty} \ \epsilon \right \rangle.
\end{align}
A straightforward application of Isserlis' Theorem \citep{Isserlis} gives that $\left \langle \delta r_{\infty}^2 \right \rangle = (1 + \rho^2)/N$, $ \left \langle \epsilon^2 \right \rangle = 4 \left(1 + \rho^2 \right)/N$, and $\left \langle \delta r_{\infty} \epsilon \right \rangle = 4 \rho/N$. Putting everything together, we obtain
\begin{align}
\lim_{N \rightarrow \infty} N\left \langle \delta r_{\rm q}^2 \right \rangle = \frac{\left(1 - \rho^2\right)^2}{1+\rho^2}.
\end{align}
We can easily verify that this result is equal to the CRB:
\begin{align}
\left \langle \delta r_{\mathrm{CR}}^2 \right \rangle &= \left\{ N \int_{-\infty}^\infty dx dy\,  \frac{ \left(\frac{\partial P(x,y;\rho)}{\partial \rho} \right)^2 }{P(x,y;\rho)}  \right\}^{-1}\\
\nonumber &= \left\{ 
 \frac{N}{\left( 1-\rho^2 \right)^4} \int_{-\infty}^\infty dx dy\, \left[ -\rho\left(1 - \rho^2 \right) - \left(1 + \rho^2 \right) x y + \rho \left( x^2 + y^2 \right) \right]^2 \frac{1}{2\pi \sqrt{1-\rho^2}} \exp \left[ -\frac{\left( x^2 + y^2 - 2\rho x y \right)}{2\left(1 - \rho^2 \right)}   \right]
\right\}^{-1}\\
\nonumber &= \frac{1}{N} \frac{ \left(1 - \rho^2 \right)^2}{1 + \rho^2}.
\end{align}

\bibliography{optimal_correlation_references.bib}

\begin{thebibliography}{18}
\expandafter\ifx\csname natexlab\endcsname\relax\def\natexlab#1{#1}\fi

\bibitem[{Anderson(1996)}]{a96}
Anderson, T.~W. 1996, Statistical Science, 11, pp. 20

\bibitem[{{Cole}(1968)}]{Cole_68}
{Cole}, T. 1968, Australian Journal of Physics, 21, 273

\bibitem[{{Cooper}(1970)}]{Cooper70}
{Cooper}, B.~F.~C. 1970, Australian Journal of Physics, 23, 521

\bibitem[{{Deller} {et~al.}(2007){Deller}, {Tingay}, {Bailes}, \&
  {West}}]{DiFX}
{Deller}, A.~T., {Tingay}, S.~J., {Bailes}, M., \& {West}, C. 2007, \pasp, 119,
  318

\bibitem[{{Dewdney} {et~al.}(2009){Dewdney}, {Hall}, {Schilizzi}, \&
  {Lazio}}]{SKA_Dewdney}
{Dewdney}, P.~E., {Hall}, P.~J., {Schilizzi}, R.~T., \& {Lazio}, T.~J.~L.~W.
  2009, IEEE Proceedings, 97, 1482

\bibitem[{Fisher(1915)}]{f1915}
Fisher, R.~A. 1915, Biometrika, 10, pp. 507

\bibitem[{Fisher(1921)}]{f1921}
---. 1921, Metron, 1, 3

\bibitem[{{Fridman}(2009)}]{fridman_robust}
{Fridman}, P. 2009, \aap, 502, 401

\bibitem[{{Gwinn}(2004)}]{gwinn04}
{Gwinn}, C. 2004, \pasp, 116, 84

\bibitem[{{Gwinn}(2006)}]{gwinn06}
{Gwinn}, C.~R. 2006, \pasp, 118, 461

\bibitem[{{Hagen} \& {Farley}(1973)}]{Hagen_Farley_73}
{Hagen}, J.~B., \& {Farley}, D.~T. 1973, Radio Science, 8, 775

\bibitem[{Hotelling(1953)}]{h53}
Hotelling, H. 1953, Journal of the Royal Statistical Society. Series B
  (Methodological), 15, pp. 193

\bibitem[{Isserlis(1918)}]{Isserlis}
Isserlis, L. 1918, Biometrika, 12, pp. 134

\bibitem[{{Jenet} \& {Anderson}(1998)}]{ja98}
{Jenet}, F.~A., \& {Anderson}, S.~B. 1998, \pasp, 110, 1467

\bibitem[{{Kendall} \& {Stuart}(1979)}]{Kendall_Stuart_v2}
{Kendall}, M., \& {Stuart}, A. 1979, {The advanced theory of statistics. Vol.2:
  Inference and relationship}

\bibitem[{Lehmann \& Casella(1998)}]{Lehmann_Casella}
Lehmann, E.~L., \& Casella, G. 1998, {Theory of Point Estimation} (New York:
  {Springer-Verlag})

\bibitem[{{Thompson} {et~al.}(2001){Thompson}, {Moran}, \& {Swenson}}]{tms}
{Thompson}, A.~R., {Moran}, J.~M., \& {Swenson}, Jr., G.~W. 2001,
  {Interferometry and Synthesis in Radio Astronomy, 2nd Edition}

\bibitem[{Van~Vleck \& Middleton(1966)}]{vvm66}
Van~Vleck, J.~H., \& Middleton, D. 1966, Proceedings of the IEEE, 54, 2

\end{thebibliography}

\end{document}